# Electron-lattice energy exchange in metal nanoparticles. Quantum-kinetic and classical approaches.

P.M. Tomchuk[a] Y. Bilotsky[b],*

[a]Institute of Physics, N.A.S. of Ukraine, UA-03028 Kyiv, Ukraine

[b]Aalto University Foundation School of Chemical Technology,
Department of Material Science and Engineering,
P.O. Box 16200, FIN-00076 AALTO, Finland
*E-mail: yevgen.bilotsky@aalto.fi

*Abstract*-**We obtained the electron-lattice energy transfer constant in metal nanoparticles (MN), in quantum-mechanical and classical approach using the deformation potential Bardeen-Shockley and found the changes of the electron-lattice energy exchange (due to the finite size MN) in the quantum kinetic approach caused by the discrete phonon spectrum. The condition when the discrete phonon spectrum could be observed via the electron-phonon energy exchange has been obtained. It was shown that the classical approach can be generalized for metallic clusters with ballistic motion of electrons (electrons move from one potential wall to another one). In this case, the electron-lattice energy exchange is quasi-periodic depending on the size of the cluster, and at certain cluster size, energy exchange is particularly size sensitive.**

I. INTRODUCTION

Metal nanoparticles (MN) and their ensembles have unique physical properties [1], [2] that distinguish them from bulk metals and from macromolecules. As an example, MN have been deposited on the dielectric surface for changing reflection [3], are used in bio-sensors [4] and in genetics to visualize cells structure [5], in medicine as an antibacterial agent [6], as well as in the treatment of cancer [7]. The crucial point in these phenomena is the appearance of hot electrons, which is associated with intensity of electron-phonon interaction. In turn, this interaction is size dependent for small metal clusters, in particular for MN. Therefore, the study of the features of this interaction is an important task.

Nowadays the whole range of metal nanoparticles, from a few atoms clusters to mega clusters, widely studied. Physical properties of small clusters close to the properties of macromolecules, and the properties of large clusters are close to properties of infinite solids. However, when the size of nanoclusters becomes comparable with the size of some physical parameters, such as the mean free path of electrons and phonons, skin-layer, etc., properties of a cluster change significantly. In this paper, we shall speak mainly about of clusters, with the size of the order to (or less than) the mean free path of electrons but large than the Debye wavelength of electrons.

The optical properties of MN has been often used owing to their ability to intensively absorb and scatter light and generate strong local field near the plasmon resonance frequencies. During the absorption of intense laser radiation, electron temperatures can reach very high level. Such electrons are called the hot electrons. Hot electrons modify the optical properties of MN which leads to some nonlinear effects.

The two-temperature approximation [8] for the hot electrons phenomena description is widely used. In this model the hot electrons and phonons characterized by their own temperatures. These temperatures are usually obtained by solving the equations of energy balance. In most publications the equations for the infinite metals have been used (as it was in the original paper [8]), even for investigation of nano clusters. Usually, these equations modified for MN by adding the surface terms only (see the review[18]). But the bulk energy exchange depends on the size of the MN. When the MN size becomes smaller than the electron mean free path then its motion becomes ballistic (electron moves free from wall to wall), and this is manifested in the energy exchange [9]. In this case the constant of electron-phonon interaction oscillates as a function of the MN size and disappear if this size is smaller than the critical size. For such size only the surface energy exchange remains [10].

In [8] has been shown (for the infinite metal) how the electron-phonon energy exchange can be described in kinetically and classical approaches. We showed in [11] how classical approaches can be generalized to the case of ballistics electron movements. Our objective in this paper is:
- Firstly, starting from deformation potential Bardeen-Shockley, obtain the bulk electron-lattice energy transfer constant in both, quantum-mechanical and classical approach;
- Secondly, to explore the changes caused by the discrete phonon spectrum to the electron-lattice energy exchange (due to the finite size MN) in the quantum kinetic approach.

## II. KINETIC APPROACH TO THE HOT ELECTRONS.

The irradiation of MN or bulk metal leads to heating of the electron gas. Due to electron-lattice interaction, the metal lattice is also heated. In the case of MN incorporated into the dielectric matrix, the electrons in MN have their own temperature and the lattice of metal cluster has its own temperature. The equation of energy balance to determine the temperatures, can be consistently obtained from the kinetic equations for distribution functions of electrons and phonons. However, there is the size of MN for which the kinetic method in a standard form is not applicable. Increasing the size of MN leads to transition to the bulk metals, where the movement of electrons of the conduction band has quasi classic character, while decreasing the size of MN leads to a quasi molecular metal clusters, in which electrons move on the quantum orbits. In this article the quasi molecular metal clusters will not be discussed. We assume that the conduction electrons in the MN can be characterized by quasi continuous momentum $\vec{p} = \hbar \vec{k}$ ($\vec{k}$ is the wave vector) and the distribution function $f_{\vec{k}}$ for probability to find the quantum state with momentum $\vec{k}$. The kinetic equation for MN being irradiated by electro magnetic field, can be written as

$$\frac{\partial f_{\vec{k}}}{\partial t} + \vec{v} \frac{\partial f_{\vec{k}}}{\partial \vec{r}} + \frac{e \vec{E}_{in}}{\hbar} \frac{\partial f_{\vec{k}}}{\partial \vec{k}} + \hat{I} f_{\vec{k}} = 0, \quad (1)$$

where $t$ - time, $\vec{r}$ is the coordinates vector of the electron, $\vec{v}$ - its speed ($\vec{v} = \hbar \vec{k}/m$, $m$ - electron's mass), $\vec{E}_{in}$ - electric field, induced by the external field of the electromagnetic waves in MN and $\hat{I} f_{\vec{k}}$ - collision integral.

The collision integral consists of the sum of two integrals

$$\hat{I} = \hat{I}_{ee} + \hat{I}_{ph}, \quad (2)$$

where $\hat{I}_{ee}$ is the integral of electron-electron collisions and $\hat{I}_{ph}$ - the integral of electron-phonon collisions.

We will consider the relaxation processes after the electron temperature (due to intense electron-electron interaction) has been set ($\tau_{ee} \sim 10^{-14} s$), therefore, in this stage, the integral of electron-electron collisions is omitted ($\hat{I}_{ee} f_{\vec{k}} \approx 0$).

The integral of electron-phonon collisions can be written in the usual form:

$$\hat{I}_{ph} f_{\vec{k}} = \sum_{\vec{k}',\vec{q}} W_{\vec{k},\vec{k}',\vec{q}} \left\{ \begin{array}{l} \left[ \left( N_{\vec{q}} + 1 \right) f_{\vec{k}} (1 - f_{\vec{k}'}) - N_{\vec{q}} f_{\vec{k}'} (1 - f_{\vec{k}}) \right] \delta \left[ \varepsilon_{\vec{k}'} - \varepsilon_{\vec{k}} + \hbar \omega_{\vec{q}} \right] + \\ + \left[ N_{\vec{q}} f_{\vec{k}} (1 - f_{\vec{k}'}) - (N_{\vec{q}} + 1) f_{\vec{k}'} (1 - f_{\vec{k}}) \right] \delta \left[ \varepsilon_{\vec{k}'} - \varepsilon_{\vec{k}} - \hbar \omega_{\vec{q}} \right] \end{array} \right\}. \quad (3)$$

Here $W_{\vec{k},\vec{k}',\vec{q}}$ the probability of electron's transition from the state $\vec{k}$ to the state $\vec{k}'$ per second, $N_{\vec{q}}$ is the distribution function of phonon's energy $\hbar \omega_{\vec{q}}$ ($\vec{q}$ is the phonon wave vector). Then we will take $N_{\vec{q}}$ as the Planck's function:

$$N_{\vec{q}} = \left\{ \exp\left(\frac{\hbar \omega_{\vec{q}}}{\theta}\right) - 1 \right\}^{-1}, \qquad (4)$$

where $\theta$ is the phonon temperature in energy units (i.e. $\theta \equiv k_B T$, $k_B$ - Boltzmann constant and $T$ is the absolute temperature).

Based on a second quantization the Hamiltonian of the electron-phonon interactions comes as

$$\hat{H}_{int} = \sum_{\vec{k},\vec{k}',\vec{q}} C_{\vec{k},\vec{k}',\vec{q}} \left\{ b_{\vec{q}} + b^+_{-\vec{q}} \right\} a^+_{\vec{k}} a_{\vec{k}'}, \qquad (5)$$

where $a^+_{\vec{k}}$ and $a_{\vec{k}}$ are creation and annihilation operators for an electron in the state $\vec{k}$ (and similarly for the phonon's operators $b^+_{\vec{q}}$ and $b_{\vec{q}}$), then according to quantum mechanics

$$W_{\vec{k},\vec{k}',\vec{q}} = \frac{2\pi}{\hbar} \left| C_{\vec{k},\vec{k}',\vec{q}} \right|^2 \qquad (6)$$

For obtaining the Hamiltonian of the electron-phonon interaction in the form (3) we use the deformation potential Bardeen-Shockley

$$H_{int} = -\Lambda \, div \, \vec{u}, \qquad (7)$$

with $\Lambda$ - the energy constant of the deformation potential and $\vec{u}$ is the displacement vector of the lattice:

$$\vec{u} = \left(\frac{\hbar}{2M_i \cdot N}\right)^{1/2} \sum_{\vec{q}} \frac{\vec{e}_{\vec{q}}}{\sqrt{\omega_{\vec{q}}}} \left\{ b_{\vec{q}} + b^+_{-\vec{q}} \right\} e^{i\vec{q}\vec{r}}, \qquad (8)$$

where $M_i$ is the ion mass, $N$ is the number of ions in MN and $\vec{e}_{\vec{q}}$ - mutually orthogonal unit vectors. Going to the second quantization of electron variables in Eq. (7) we have:

$$\hat{H}_{int} = \int d\vec{r} \, \Psi^+(\vec{r}) \, H_{int} \, \Psi(\vec{r}),$$
$$\Psi(\vec{r}) = \frac{1}{\sqrt{V}} \sum_{\vec{k}} a_{\vec{k}} \, e^{i\vec{k}\vec{r}}, \qquad (9)$$

$V$ - volume of MN.

From the expressions Eqs. (7), (8) and (9) the formula (6) comes as:

$$W_{\vec{k},\vec{k}',\vec{q}} = \frac{\pi \Lambda^2}{\rho s V} q \, \delta_{\vec{k}',\vec{k}-\vec{q}}, \qquad (10)$$

here we used the relation $M_i \cdot N = \rho V$ with $\rho$ - density, and the dispersion equation $\omega_{\vec{q}} = s q$ for longitudinal phonons (the only kind of phonons which electrons interacted with), and $s$ is the phonon's speed. Next, we assume that under the laser irradiation of MN the distribution function of electrons can be written as:

$$f_{\vec{k}} = f_0(\varepsilon_{\vec{k}}) + f^{(1)}_{\vec{k}}, \qquad (11)$$

where $f_0(\varepsilon_{\vec{k}}) = \dfrac{1}{e^{\frac{\varepsilon_{\vec{k}} - \mu}{\theta_e}} + 1}$ is the Fermi distribution function with the effective temperature $\theta_e$. The small correction $f^{(1)}_{\vec{k}}$ to the Fermi distribution, is the iterative solution of the kinetic equation (1).

Multiplying the equation (1) by the $\varepsilon_{\vec{k}}$ and summing (or integrate) over all $\vec{k}$ we receive the equation for the electrons effective temperature (the energy balance equation)

$$\frac{\partial}{\partial t}\left(C_e \, T_e\right) = div\left(K_e \vec{\nabla} T_e\right) + Q - \left(\frac{\partial \tilde{\varepsilon}}{\partial t}\right)_{e,ph} \qquad (12)$$

where $C_e$ is the heat capacity and $K_e$ - the thermal conductivity of the electron gas. The energy absorbed of MN per unit volume is $Q = \vec{E}_{in} \vec{j}$ (with current density $\vec{j} = \frac{1}{\hbar} \sum_{\vec{k}} \frac{\partial \varepsilon_{\vec{k}}}{\partial \vec{k}} f_{\vec{k}}^{(1)}$ caused by the field $\vec{E}_{in}$). The expression of the electron-phonon energy exchange (12) converts (by using Eqs. (3) and (4)) to:

$$\left(\frac{\partial \tilde{\varepsilon}}{\partial t}\right)_{e,ph} = \sum_{\vec{k}} \varepsilon_{\vec{k}} \hat{I} f_0(\varepsilon_{\vec{k}}) =$$
$$= \sum_{\vec{k},\vec{k}',\vec{q}} W_{\vec{k},\vec{k}',\vec{q}} N_{\vec{q}} \hbar \omega_{\vec{q}} \left\{ e^{\frac{\hbar \omega_{\vec{q}}}{\theta}} f_0(\varepsilon_{\vec{k}})(1 - f_0(\varepsilon_{\vec{k}'})) - f_0(\varepsilon_{\vec{k}'})(1 - f_0(\varepsilon_{\vec{k}})) \right\} \times \quad (13)$$
$$\times \delta(\varepsilon_{\vec{k}'} - \varepsilon_{\vec{k}} + \hbar \omega_{\vec{q}})$$

Assume that the lattice and electrons temperature satisfy the inequality $\hbar \omega_{\vec{q}} \ll \theta_e, \theta$. Expanding all functions (except $\delta$ - function) as power series of $\hbar \omega_{\vec{q}}$ and using the relation $f_0(\varepsilon_{\vec{k}'}) = f_0(\varepsilon_{\vec{k}} - \hbar \omega_{\vec{q}})$, we received from Eq. (13):

$$\left(\frac{\partial \tilde{\varepsilon}}{\partial t}\right)_{e,ph} \approx \sum_{\vec{k},\vec{k}',\vec{q}} W_{\vec{k},\vec{k}',\vec{q}} \left(\frac{\theta}{\hbar \omega_{\vec{q}}}\right) \cdot (\hbar \omega_{\vec{q}})^2 \left\{ \frac{f_0(\varepsilon_{\vec{k}})(1 - f_0(\varepsilon_{\vec{k}}))}{\theta} + \frac{\partial f_0(\varepsilon_{\vec{k}})}{\partial \varepsilon_{\vec{k}}} \right\} \cdot \delta(\varepsilon_{\vec{k}'} - \varepsilon_{\vec{k}} + \hbar \omega_{\vec{q}}).$$
(14)

Then take into account the expression (10) and the identity

$$f_0(\varepsilon_{\vec{k}})(1 - f_0(\varepsilon_{\vec{k}})) = -\theta_e \frac{df_0(\varepsilon_{\vec{k}})}{d\varepsilon_{\vec{k}}}$$

we receive:

$$\left(\frac{\partial \tilde{\varepsilon}}{\partial t}\right)_{e,ph} = \frac{\pi \hbar \Lambda^2}{V \rho} \sum_{\vec{k},\vec{q}} q^2 (\theta_e - \theta) \left(-\frac{\partial f_0(\varepsilon_{\vec{k}})}{\partial \varepsilon_{\vec{k}}}\right) \delta(\varepsilon_{\vec{k}-\vec{q}} - \varepsilon_{\vec{k}} + \hbar \omega_{\vec{q}}). \quad (15)$$

Until now we did not explore the condition when the domain is bounded, and thence phonons and electrons spectrums is discrete.
Before exploring the finite size effects, consider the infinite metals. In this case we can replace the sums by the integrals in Eq. (15)

$$\sum_{\vec{q}} \to \frac{V}{(2\pi)^3} \int d\vec{q} \quad , \quad \sum_{\vec{k}} \to \frac{2V}{(2\pi)^3} \int d\vec{k}.$$

Ignoring $\hbar \omega_{\vec{q}}$ in the argument of $\delta$ -function in Eq. (15) (elastic approximation) we come to the expression:

$$\sum_{\vec{q}} q^2 \delta(\varepsilon_{\vec{k}-\vec{q}} - \varepsilon_{\vec{k}}) = \frac{V}{(2\pi)^3} \int d\vec{q} \, q^2 \, \delta(\varepsilon_{\vec{k}-\vec{q}} - \varepsilon_{\vec{k}}) = \frac{V}{(2\pi)^3} \frac{2\pi m}{\hbar^2 k} \frac{q_{max}^4}{4} \quad (16)$$

with

$$q_{max} = \begin{cases} q_D, & \text{if } q_D \leq 2k_F \\ 2k_F, & \text{if } q_D > 2k_F \end{cases}, \quad (17)$$

where $q_D$ is the wave vector corresponding to the Debye frequency and $k_F$ is the Fermi wave vector of the electron. On applying well knowing approximation

$$-\frac{\partial}{\partial \varepsilon_{\vec{k}}} f_0(\varepsilon_{\vec{k}}) \approx \delta\left(\varepsilon_{\vec{k}} - \mu\right) \tag{18}$$

and using the expression (16) we transfer Eq. (15) into the form

$$\left(\frac{\partial \tilde{\varepsilon}}{\partial t}\right)_{e,ph} = (\theta_e - \theta) \frac{m^2 \Lambda^2}{2\hbar^3} \frac{V}{(2\pi)^3 \rho} \begin{Bmatrix} q_D^4; & q_D \leq 2k_F \\ (2k_F)^4, & q_D > 2k_F \end{Bmatrix}. \tag{19}$$

This equation coincides with Eq. (9) of [8] if $q_{max} = q_D$ and $V = 1$, with $q_D = k_\beta T_D / \hbar s$, where is the Debye temperature $T_D$. The equation for the electron-lattice energy exchange (19) is valid for massive metal for temperatures $\theta > \hbar \omega_D$.

In the next section we will show how the classic approach for expression of the intensity of the electron-lattice energy exchange can be relatively easily generalized to the case, when the motion of electrons is not in diffuse regime but in the ballistic one (when the cluster size becomes smaller than the mean free path of an electron). It worth to notice that in this regime we can't use the equation the quantum kinetic (3) in present form because we explore the approximation

$$\left|\frac{e^{\frac{i}{\hbar}\Delta\varepsilon t} - 1}{\Delta \varepsilon / \hbar}\right|^2 = \lim_{t \to \infty} \left(\frac{\sin\left(\frac{\Delta \varepsilon t}{2\hbar}\right)}{\Delta \varepsilon / 2\hbar}\right)^2 = 2\pi \hbar t \delta(\Delta \varepsilon), \tag{20}$$

$$\Delta \varepsilon = \varepsilon_{\vec{k} \pm \vec{q}} - \varepsilon_{\vec{k}} \pm \hbar \omega_{\vec{q}} \tag{21}$$

to deliver the collision integral in Eq. (3). The limit transition (20) is valid if

$$\frac{\Delta \varepsilon}{\hbar} \tau \gg 1, \tag{22}$$

where $\tau$ is the electron relaxation time. The feasibility to use this delta function in Eq. (3) allows us to speak about the law of conservation of energy (in the energy exchange between electrons and phonons). In this case, the electron can transfer the energy only less or equal to the maximum phonon energy – Debye energy. But, when the size of nanoparticles is smaller then the mean free path of an electron, one shall use the electron passage time $\tau_{ballistic} \simeq \frac{L}{v} \approx \frac{L}{v_F}$ (from wall to wall) instead of $\tau$ in the inequality (22). While the cluster size is decreasing, the ratio

$$\max \frac{\Delta \varepsilon}{\hbar} \tau = \frac{\max \Delta \varepsilon}{\hbar} \tau_{ballistic} \simeq \frac{\hbar \omega_D}{\hbar} \frac{L}{v_F} \approx 1 \tag{23}$$

can be achieved. This means that Eq. (20) not valid anymore and we can't use delta function in Eq. (3). Therefore, there are clusters sizes for which the quantum kinetic equation can not be applied in its conventional form. To obtain the integral electron-phonon collisions which would adequately describe the quasi-ballistic regime of electron motion in a MN, the quasi-classic wave functions of electrons should be used. But this is a separate problem and we will not deal with this now.

III. CLASSICAL DESCRIPTION OF ELECTRON-LATTICE ENERGY EXCHANGE IN CONFINED SPACE.

In the previous section we obtained the expression for the electron-lattice energy exchange (19) in the kinetic equation approach. In this section we solve the same problem in the classical approach. The classic approach proposed in [12], [8] for infinite metal and has been generalized in our publications [11], [13], [9] for MN with ballistic electrons motions. Electrons in the MN are ballistic if the size of MN is smaller then the mean free electron path. Comparing the results of kinetic and classical approaches, we can trace, in which conditions the results coincide, and that brings what is new in the energy exchange for the ballistic motion of electrons (from one potential wall to the opposite one). In the classical description of electron-lattice energy exchange we start from the equation for the lattice displacement vector associated with the longitudinal acoustic lattice vibrations generated by moving electrons [13]

$$\frac{\partial^2}{\partial t^2}\vec{u} - s^2 \Delta \vec{u} = -\frac{\Lambda}{\rho}\vec{\nabla}\delta(\vec{r}_0(t) - \vec{R}) \qquad (24)$$

where $\vec{r}_0(t)$ is the electron trajectory. It is convenient to use scalar $\chi$ instead of vector $\vec{u}$

$$\vec{u} = \vec{\nabla}\chi. \qquad (25)$$

From Eq. (24) we obtain the equation for $\chi$:

$$\frac{\partial^2}{\partial t^2}\chi - s^2 \Delta \chi = -\frac{\Lambda}{\rho}\delta(\vec{r}_0(t) - \vec{R}). \qquad (26)$$

Since the right side of Eq. (20) is the force that generates the longitudinal acoustic lattice vibrations, then the energy transferring from an electron to the lattice per seconds is:

$$\frac{\partial \xi}{\partial t} = \Lambda \int \frac{\partial \vec{u}}{\partial t}\vec{\nabla}\delta(\vec{R}-\vec{r}_0(t))d^3R = \\ = \Lambda \int \left(\vec{\nabla}\frac{\partial \chi}{\partial t}\right)\vec{\nabla}\delta(\vec{R}-\vec{r}_0(t))d^3R. \qquad (27)$$

The solution of Eq. (26) gives (via Eq. (27)) the energy transmitted to the lattice. We assume that all trajectories of electrons which pass near periodic orbit between potential walls (with the distance between them $L$) will remain near it. Choose $Z$ axis along the velocity vector $\vec{v}$. Then the trajectory of an electron can be written as:

$$\vec{r}_0(t) = \{0,0,z_0(t)\} \\ z_0(t) = \begin{cases} vt, & t \leq \tau/2 \\ L - v(t-\tau/2), & t > \tau/2 \end{cases}, \qquad (28)$$

where $\tau = 2L/v$ is the period of electron's movement. Therefore we shall try to find the solution of Eq. (26) in Fourier series form:

$$\chi(\vec{R},t) = \chi(\vec{R}_\perp, R_l; t) = \sum_{l=-\infty}^{\infty}\int d\vec{k}_\perp \tilde{\chi}(\vec{k}_\perp, k_l; t)\exp\{i(\vec{k}_\perp \vec{R}_\perp + k_l R_l)\} \qquad (29)$$

where

$$k_l = l\frac{2\pi}{L} \qquad (30)$$

and periodic condition

$$\chi(\vec{R}_\perp, R_l; t) = \chi(\vec{R}_\perp, R_l + L; t) \qquad (31)$$

has been applied. In this Section we use $\vec{k}$ as the wave vector of the Fourier expansion (unlike the previous Section, where $\vec{k}$ is the electron wave vector). Upon inserting Eq. (29) into Eq. (26) we obtain:

$$\tilde{\chi}(\vec{k}_\perp, k_l; t) = -\frac{\Lambda}{(2\pi)^2 L}\frac{\exp(-i\vec{k}\vec{v}_\tau t)}{(\vec{k}\vec{v})^2 - (ks)^2}. \qquad (32)$$

In (32) we have denoted

$$\vec{v}_\tau = \begin{cases} \vec{v}, & t < \tau/2 \\ -\vec{v}, & t > \tau/2 \end{cases} \qquad (33)$$

and $\vec{k} = \{\vec{k}_\perp, k_l\}$, $\vec{k}\vec{v} - k_l v$.

The expression (28) has a pole and therefore the integral (23) becomes undefined. Therefore, we proceed similarly to [8], assuming that the speed of sound has a small imaginary term $s = s_0 + i\varepsilon$ which is responsible for weak sound damping.

The corresponding integral has to be taken in the sense of principal value by using the formula

$$\lim_{\varepsilon \to 0} \frac{1}{(k_l \upsilon)^2 - k^2(s_0 + i\varepsilon)^2} \to P \frac{1}{(k_l \upsilon)^2 - (k s_0)^2} + i\pi \delta\{(k_l \upsilon)^2 - (k s_0)^2\}. \tag{34}$$

Note that in Eq. (29) we used the complex variables but only the real part of Eq. (27) $\text{Re}\left(\frac{\partial \xi}{\partial t}\right)$ has physical meaning. By substituting of Eq. (32) into Eq. (27) and applying the formula (34) with the expression $\exp\{i k_l z_0(t)\} = \exp\{i k_l \upsilon_\tau t\}$, we get

$$\text{Re}\left(\frac{\partial \xi}{\partial t}\right) = \frac{\Lambda^2 \upsilon}{\rho L} \sum_{l=1}^{\infty} k_l \int_0^{k_{max}} dk_\perp k_\perp (k_\perp^2 + k_l^2) \delta\{(k_l \upsilon)^2 - (k s_0)^2\}, \tag{35}$$

where $k_{max}$ is the maximum of $k_\perp$ which corresponds to the Debye wave vector $q_D$ ($k_{max} = q_D = \frac{\omega_D}{s_0}$).

After calculation of the integral in Eq. (35) we receive (see the details in [13]);

$$\text{Re}\left(\frac{\partial \xi}{\partial t}\right) = \frac{\Lambda^2}{2\rho L \upsilon} \left(\frac{\upsilon}{s_0}\right)^4 \sum_{l=1}^{l_{max}} k_l^3 \tag{36}$$

The maximum value of $l_{max}$ is determined from the condition that the argument of the $\delta$ function is zero if $k_\perp \leq k_{max}$, i.e.

$$\left(\frac{\upsilon^2}{s_0^2} - 1\right) k_l^2 \leq k_{max}^2. \tag{37}$$

This condition and Eq. (30) gives

$$l_{max} \leq \frac{L}{2\pi} \frac{k_{max}}{(\upsilon^2/s_0^2 - 1)^{1/2}} \approx L \frac{k_{max}}{2\pi} \frac{s_0}{\upsilon} \equiv \eta. \tag{38}$$

The inequality $l_{max} < 1$ means that the argument of the $\delta$ function has no zeros. From this it follows that $\text{Re}\left(\frac{\partial \xi}{\partial t}\right) = 0$ there will be no energy exchange between electrons and electrons, except surface scattering.

Let us to define the function $\text{floor} \, \eta \equiv \lfloor \eta \rfloor$, which returns the largest integer not greater than $\eta$. Using this function we can rewrite Eq. (38) as

$$l_{max} \approx \left\lfloor L \frac{k_{max}}{2\pi} \frac{s_0}{\upsilon} \right\rfloor = \lfloor \eta \rfloor. \tag{39}$$

The sum in Eq. (36) can be calculated easily

$$\sum_{l=1}^{l_{max}} k_l^3 \xrightarrow{k_l = l \frac{2\pi}{L}} \left(\frac{2\pi}{L}\right)^3 \sum_{l=1}^{l_{max}} l^3 = \left(\frac{2\pi}{L}\right)^3 \frac{l_{max}^4}{4} \left(1 + \frac{1}{l_{max}}\right)^2. \tag{40}$$

Therefore the expression (36) comes as

$$\text{Re}\left(\frac{\partial \xi}{\partial t}\right) = \frac{\Lambda^2 k_{max}^4}{16\pi \rho \upsilon} q(\eta), \tag{41}$$

with

$$q(\eta) = \begin{cases} \left(\frac{\lfloor \eta \rfloor}{\eta}\right)^4 \left(1 + \frac{1}{\lfloor \eta \rfloor}\right)^2, & \lfloor \eta \rfloor \geq 1 \\ 0, & \lfloor \eta \rfloor < 1 \end{cases}. \tag{42}$$

The result of $\text{Re}\left(\frac{\partial \xi}{\partial t}\right)$ for infinite metal (see [8]) comes from Eq. (41),

$$\text{Re}\left(\frac{\partial \xi}{\partial t}\right)_\infty \equiv \lim_{L\to\infty} \text{Re}\left(\frac{\partial \xi}{\partial t}\right) = \frac{\Lambda^2 k_{max}^4}{16\pi \rho \upsilon} \lim_{L\to\infty} q(L,\upsilon) = \frac{\Lambda^2 k_{max}^4}{16\pi \rho \upsilon}$$

because $\lim_{L\to\infty} q(\eta) = 1$.

At finite $L$ the function $q(\eta)$ has quasi-oscillating dependence on $L/L_c$. In Figure 1 shows the dependence of $q(x)$ on the dimensionless argument $x = L/L_c$, where

$$L_c = \frac{2\pi \upsilon}{k_{max} s_0} = \frac{2\pi \upsilon}{\omega_D}$$

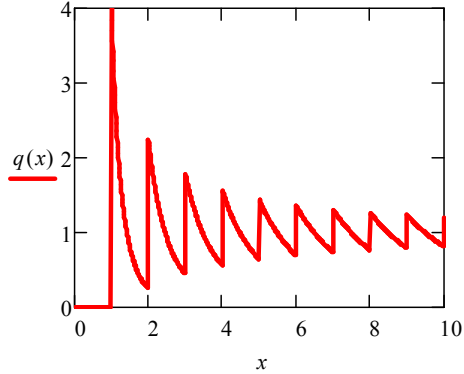

Fig. 1 The chart shows the dependence of $q(x)$ on the dimensionless parameter $x = L/L_c$.

The emergence of each new peak of the curve occurs due to the inclusion of the new acoustic mode in the electron-lattice energy exchange. The expression (41) determines the energy that an electron moving at a speed of $\upsilon$ transfers to lattice per unit time. The contribution of all electrons per unit volume of the MN to the electron-lattice energy exchange, which allowed by the Paul principle, comes as

$$\left(\frac{\partial \tilde{\varepsilon}}{\partial t}\right)_{e,ph} = \int_{\mu-\theta_e}^{\mu+\theta_e} d\varepsilon \left\{ 2g(\varepsilon) f_0(\varepsilon) \text{Re}\frac{\partial \xi}{\partial t} \right\}. \tag{43}$$

where

$$g(\varepsilon) = \frac{(2m)^{3/2}}{(2\pi)^2 \hbar^3} \sqrt{\varepsilon} \tag{44}$$

is the density of states and the spin of an electron gives the factor 2.
In the case of infinite metal, according to Eq. (41)

$$\lim_{L\to\infty} \text{Re}\left(\frac{\partial \xi}{\partial t}\right) = \frac{\Lambda^2 q_D^4}{16\pi \rho \upsilon} \tag{45}$$

and then from Eqs. (43) and (45) we get:

$$\left(\frac{\partial \tilde{\varepsilon}}{\partial t}\right)_{e,ph} \approx \theta_e \frac{m^2 \Lambda^2}{2(2\pi \hbar)^3} \frac{q_D^4}{\rho}. \tag{46}$$

The expression (42) determines the hot electrons energy losses per second through generating of the lattice acoustic vibrations. But apart from excitation of acoustic waves, electrons also absorb acoustic energy. In the thermodynamic equilibrium state, the energy of sound generation by electrons is equal to the energy of acoustic waves absorbed by electrons. Therefore, taking into account both effects (generation and absorption) in Eq. (46) instead of $\theta_e$ only we shall write $\theta_e - \theta$ and that giving the result

$$\left(\frac{\partial \tilde{\varepsilon}}{\partial t}\right)_{e,ph} \approx (\theta_e - \theta)\frac{m^2 \Lambda^2}{2(2\pi \hbar)^3}\frac{q_D^4}{\rho}. \tag{47}$$

Comparing Eq. (47) with Eq. (19) (if $q_D < 2k_F$) we see that for infinite metals, the same results come as from classical approach and from quantum-kinetic as well (for $\theta \gg \hbar\omega_D$).

In the case of finite size MN, where integer $\eta(L,\upsilon)$ does not falls into the range $\mu - \theta_e \leq \varepsilon \leq \mu + \theta_e$, then Eq. (47) yields:

$$\left(\frac{\partial \tilde{\varepsilon}}{\partial t}\right)_{e,ph} \approx (\theta_e - \theta)\frac{m^2 \Lambda^2}{2(2\pi \hbar)^3}\frac{q_D^4 \cdot q(L,\upsilon_F)}{\rho}. \tag{48}$$

In the general case (for arbitrary $L$) instead of Eq. (48) we have:

$$\left(\frac{\partial \tilde{\varepsilon}}{\partial t}\right)_{e,ph} = (\theta_e - \theta)\frac{m^2 \Lambda^2}{2(2\pi \hbar)^3}\frac{q_D^2}{\rho}\int_{-1}^{1} du\, f_0(u)\, \tilde{q}(L,u), \tag{49}$$

where $u = \dfrac{\varepsilon - \mu}{\theta_e}$. We obtained $\tilde{q}(L,u)$ from $L(q,\upsilon)$ by of the substitution

$$\upsilon \to \upsilon_F \left(1 + \frac{\theta_e}{\mu}u\right)^{1/2}.$$

It should be notice that for the fixed value of $L$ the peak function $\tilde{q}(L,u)$ which falls into the interval $-1 \leq u \leq 1$, the integral $\int_{-1}^{1} du\, f_0(u)\tilde{q}(L,u)$ is very sensitive to the location of the peak. We emphasize that the appearance of peaks associated with the amount of the acoustic modes involved in energy exchange.

### IV. MANIFESTATION OF THE DISCRETE PHONONS SPECTRUM IN THE ELECTRON-LATTICE. ENERGY EXCHANGE

In the first Section, we have received an expression for the electron-lattice energy transfer in massive metal. This expression remains unchanged in MN which size is larger than the electron mean free path. In the Second section, we investigated the electron-lattice energy exchange in the classical approach. We have seen that for ballistic electron motion mode and for the discrete spectrum of acoustic lattice vibrations, the quasi-periodic size-dependence of electron-lattice energy transfer appear. However, there is a wide range of MN sizes and temperatures in which the quantum approach for describing the discrete phonon spectrum is requires. The electron energy spectrum can still be regarded as continuous (MN sizes much larger than the de Broglie wavelength) but discrete phonon spectrum is becoming required. In this range of MN sizes it is convenient (for describing the energy transfer) to use the phonon-electron collision integral rather than electron-lattice collision integral. This advantage emerge because the summation over electrons quasi-impulse can be replaced, for continuum electron spectrum, by the integral and this integral has been calculated explicitly. This assumption simplifies the expression for the electronic-lattice energy essentially.

It is necessary to emphasize that the MN size reduction has several consequences for the physics of hot electrons. Firstly, with decreasing the size of MN their phonon spectrum changing. In particular, the discreteness of the wave vector is essential, and the spectrum of this vector is limited not only from above, but also from below. The minimum value of the wave vector determined by the size of the MN. In addition, the surface phonons give their own contribution to the electron-phonon energy exchange ( see for example [14], [10]).

Second, depending on the characteristics of the electron-phonon energy exchange and intensity of heat removal from the MN into a matrix, the phonons temperature in the MN may be significantly differ from the temperature of the matrix. Moreover, there are situations where two-temperature approximation becomes invalid (see [15]).

Regarding the contribution of surface phonons in the energy, we already investigated this issue in [10], [16].

Here we want to investigate, using a simple model, the consequences for the bulk electron-phonon energy exchange, caused by phonons spectrum modification in MN.
Therefore, we consider that the phonons can be described by Planck equilibrium distribution function with a temperature of phonons, which may differ from the electron temperature.
Thus, the phonon-electron collision integral, when the electron-phonon interaction is characterized by deformation potential Bardeen-Shockley (see [10])), can be written as

$$\left(\frac{\partial N_{\vec{q}}}{\partial t}\right)_{st} = \frac{\pi \Lambda^2}{\rho s V} \sum_{\vec{k}} q \left\{ \begin{array}{l} (N_{\vec{q}}+1) f_{\vec{k}+\vec{q}}(1-f_{\vec{k}}) \delta[\varepsilon_{\vec{k}+\vec{q}} - \varepsilon_{\vec{k}} - \hbar \omega_{\vec{q}}] - \\ -N_{\vec{q}} f_{\vec{k}}(1-f_{\vec{k}+\vec{q}}) \delta[\varepsilon_{\vec{k}} - \varepsilon_{\vec{k}+\vec{q}} + \hbar \omega_{\vec{q}}] \end{array} \right\} = \qquad (50)$$

$$= \frac{\pi \Lambda^2}{\rho s V} \sum_{\vec{k}} q \{(N_{\vec{q}}+1) f_{\vec{k}+\vec{q}}(1-f_{\vec{k}}) - N_{\vec{q}} f_{\vec{k}}(1-f_{\vec{k}+\vec{q}})\} \delta[\varepsilon_{\vec{k}+\vec{q}} - \varepsilon_{\vec{k}} - \hbar \omega_{\vec{q}}]$$

Because we consider continuous electron spectrum, then we can replace, as we mentioned before the sum in Eq. (45) by the integral. If we take $N(\theta,q)$ as a Planck function with lattice temperature $\theta$ (formula (4)) on the right side of Eq. (50), as well as $f_0(\varepsilon_{\vec{k}})$ - the Fermi function of electrons with temperature $\theta_e$ and use the identity

$$\{f_{\vec{k}+\vec{q}}(1-f_{\vec{k}})(N_{\vec{q}}+1) - f_{\vec{k}}(1-f_{\vec{k}+\vec{q}})\} \delta[\varepsilon_{\vec{k}+\vec{q}} - \varepsilon_{\vec{k}} - \hbar \omega_{\vec{q}}] \to$$

$$\to N(\theta,\vec{q}) f_0(\varepsilon_{\vec{k}}) f_0(\varepsilon_{\vec{k}+\vec{q}}) \exp\left(\frac{\varepsilon_{\vec{k}} - \mu}{\theta_e}\right) \{e^{\hbar \omega_{\vec{q}}/\theta} - e^{\hbar \omega_{\vec{q}}/\theta_e}\} \delta[\varepsilon_{\vec{k}+\vec{q}} - \varepsilon_{\vec{k}} - \hbar \omega_{\vec{q}}] \qquad (51)$$

we get:

$$\left(\frac{\partial N_{\vec{q}}}{\partial t}\right)_{st} = \frac{m^2 \Lambda^2 \theta_e}{2 \pi \rho s \hbar^4} \{N(\theta_e,\vec{q}) - N(\theta,\vec{q})\} \ln\left\{1 + \frac{f_0(\varepsilon_{\vec{q}})}{N(\theta_e,\vec{q})}\right\}, \qquad (52)$$

where $N(\theta_e,\vec{q})$ is the Planck function, in which the temperature of the lattice $\theta$ is replaced by the temperature of the hot electrons $\theta_e$ and the energy

$$\varepsilon_q = \frac{1}{2m}\left(\frac{\hbar q}{2} + m s\right)^2. \qquad (53)$$

Before analyze Eq. (52) we note that the relation between the modules of the Fermi wave vector $k_F$ and wave vector Debye $q_D$ in metals and semimetals can be of two types:

a) $k_F \leq \frac{q_D}{2}$, b) $k_F > \frac{q_D}{2}$.

Case a) occurs in metals with low electron concentration (semimetals) and case b) in good metals with a high concentration of electrons.
Using Eq. (52) one can determines the electron-lattice energy exchange for
discrete and continuous phonon spectrum:

$$\left(\frac{\partial \varepsilon}{\partial t}\right)_{ph,e} = \sum_{\vec{q}} \hbar \omega_{\vec{q}} \left(\frac{\partial N_{\vec{q}}}{\partial t}\right)_{st}. \qquad (54)$$

The maximum value of $q$ in Eq. (54) defined by the condition $q_{max} = \min(q_D, 2k_F)$.
Consider first the case of high temperatures, when

$$\theta_e, \theta \gg \hbar q_D s. \qquad (55)$$

Then $N(\theta_e,\vec{q}) \approx \theta_e/\hbar q_D s \gg 1$ and we have that $f_0(\varepsilon_{\vec{k}})/N(\theta_e,\vec{q}) \ll 1$. Thus from (52) and (54) yields:

$$\left(\frac{\partial \varepsilon}{\partial t}\right)_{ph,e} \approx \frac{m^2 \Lambda^2}{2 \pi \rho \hbar^3} \sum_{\vec{q}} q f_0(\varepsilon_{\vec{k}})[\theta_e - \theta]. \qquad (56)$$

After substitution the summation by the integration, all calculations can be brought to an end and we get already known result - Eq. (19). Now we get it by using the phonon-electron collision integral (45) and Eq. (19) was obtained by electron-phonon collision integral. Both results are the same, as that should be. As can be seen in high-temperature regime, one can get the expression for the electron-lattice energy transfer for materials in which as the inequality a) as well as inequality b) is hold. The following discussion focuses on the metallic particles (an inequality b)). If the inequality b) is holds, then in a strongly degenerate case $(\mu \gg \theta_e)$, the simplification $f_0(\varepsilon_{\vec{k}}) \approx 1$ can be used. Then

$$\ln\left\{1 + \frac{f_0(\varepsilon_{\vec{k}}))}{N(\theta_e, \vec{q})}\right\} \approx \ln\left\{1 + \frac{1}{N(\theta_e, \vec{q})}\right\} = \hbar \omega_q / \theta_e \text{ and we have from Eq. (52):}$$

$$\left(\frac{\partial N_{\vec{q}}}{\partial t}\right)_{st} = \frac{m^2 \Lambda^2}{2\pi \rho s \hbar^4} \hbar \omega_q \left(N(\theta_e, \vec{q}) - N(\theta, \vec{q})\right) \tag{57}$$

This formula, in this form, was first obtained in [8].
Equation (57) is valid for both discrete and continuous phonon spectrum. So, according to Eqs. (54) and (57), if $\frac{q_D}{2} < k_F$, we have:

$$\left(\frac{\partial \varepsilon}{\partial t}\right)_{ph,e} \approx \frac{m^2 \Lambda^2 \theta^2}{2\pi \rho s \hbar^4} \sum_{\vec{q}} \left(\frac{qs}{\theta}\right)^2 [N(\theta_e, \vec{q}) - N(\theta, \vec{q})]. \tag{58}$$

Summation over the wave vectors $\vec{q}$ is performed by using the equality $q_i = \frac{2\pi}{L_i} n_i$. The set, which includes these numbers $n_i$ depend on the type of boundary conditions - periodic, Dirichlet or Neumann. In our model we have chosen periodic boundary conditions, for which $n_i = 0, \pm 1, \pm 2, \ldots$ and Eq. (58) is written as

$$\left(\frac{\partial \varepsilon}{\partial t}\right)_{ph,e} = \frac{m^2 \Lambda^2 \theta^2}{2\pi \rho s \hbar^6} \times \sum_{n_x = -\max n_x}^{\max n_x} \sum_{n_y = -\max n_y}^{\max n_y} \sum_{n_z = -\max n_z}^{\max n_z} \left(q \frac{\hbar s}{\theta}\right)^2 [N(\theta_e, \vec{q}) - N(\theta, \vec{q})],$$

$$q = 2\pi \sqrt{\frac{n_x^2}{L_x^2} + \frac{n_y^2}{L_y^2} + \frac{n_z^2}{L_z^2}}. \tag{59}$$

Further consider $\left(\frac{\partial \varepsilon_s}{\partial t}\right)_{ph,e}$ - the rate of change of the electron-phonon energy exchange per unit volume of a cube

$$\left(\frac{\partial \varepsilon_s}{\partial t}\right)_{ph,e} = \frac{1}{L^3} \frac{m^2 \Lambda^2 \theta^2}{2\pi \rho s \hbar^6} \times \sum_{n_x=-n}^{n} \sum_{n_y=-n}^{n} \sum_{n_z=-n}^{n} \left(q \frac{\hbar s}{\theta}\right)^2 [N(\theta_e, \vec{q}) - N(\theta, \vec{q})], \tag{60}$$

$$L \equiv L_x = L_y = L_z, \quad n \equiv \max n_x = \max n_y = \max n_z.$$

Calculation of the triple sum in Eq. (60) can be done numerically, but with increasing size of the system, computing volume is growing rapidly, so we use an analytical approach for effectively calculation these sums. To do this, we use the Walfisz-Poisson summation formula for three dimensional sums [14], [17]

$$\sum_{n_x=-\infty}^{\infty} \sum_{n_y=-\infty}^{\infty} \sum_{n_z=-\infty}^{\infty} \Psi(|\vec{n}|) = \frac{4\pi}{v} \int_{\min q_z}^{\max q_z} \Psi(q) \cdot \left[1 + \frac{1}{2\pi q} \sum_{\gamma}' \frac{\sin(2\pi \gamma q)}{\gamma}\right] \cdot q^2 dq. \tag{61}$$

In this formula $v$ is the volume per lattice point and

$$\sum_{\substack{\vec{\gamma} \in \inf, \\ |\vec{\gamma}| \neq 0}} \frac{\sin(\gamma \cdot q)}{\gamma} = 8 \sum_{j_x=1}^{\infty} \sum_{j_y=1}^{\infty} \sum_{j_z=1}^{\infty} \frac{\sin\left(L\sqrt{j_x^2 + j_y^2 + j_z^2} \cdot q\right)}{L\sqrt{j_x^2 + j_y^2 + j_z^2} \cdot q} \tag{62}$$

Equation (60) can be generalized to the case where the summation on the left side of this formula is performed by a finite set of integers, and then the sums are written as

$$\sum_{n_x=-n}^{n} \sum_{n_y=-n}^{n} \sum_{n_z=-n}^{n} q^2 \left[ N(\theta_e, \vec{q}) - N(\theta, \vec{q}) \right] =$$

$$= \frac{4\pi}{v} \int_{\min q_z}^{\max q_z} \left[ N(\theta_e, \vec{q}) - N(\theta, \vec{q}) \right] \cdot \left[ 1 + \frac{1}{2\pi q} \sum_{\gamma}{}' \frac{\sin(2\pi \gamma q)}{\gamma} \right] \cdot q^4 dq. \quad (63)$$

Let introduce in Eq. (63) the new independent variable (similar to (Kaganov M. I., Lifshitz I. M. 1957))

$$x \equiv \frac{\hbar q s}{\theta} \quad (64)$$

with some caution - in our case, the transition from the Cartesian coordinates to spherical requires explanation. In Cartesian coordinates, the maximum phonon components of the wave vector is

$$q_l = \frac{2\pi}{L_l} = \frac{2\pi}{d \cdot N_l} \quad (65)$$

where $N_l$ is the number of atomic layers being concluded along the axis. In our case we have assumed that the cubic unit cell and crystal edges are located along the main axis of the unit cell. In the spherical coordinate system there is no such a simple accordance (maximum wave vector depends not only on the wave vector of the module, but also on the angles. However, for not very small particles, for which the bulk contribution dominates and surface effects (as well as the form of particles) can be neglected, therefore

$$\left( \frac{\partial \varepsilon_s}{\partial t} \right)_{ph,e} = \frac{2}{(2\pi)^3} \frac{m^2 \Lambda^2 \theta_D^5}{\hbar^7 \rho s^4} \times \theta_e$$

$$\times \left\{ \begin{array}{l} \left[ \left( \frac{\theta_e}{\theta_D} \right) \int_{\frac{\theta_D}{\theta_e \cdot N_l}}^{\frac{\theta_D}{\theta_e}} \frac{x^4}{e^x - 1} \left\{ 1 + \frac{1}{2\pi \cdot \left( \frac{\theta_e N_l}{\theta_D} \right) x} S(x, \theta_e, \theta_D, N_l) \right\} - \\ - \left( \frac{\theta}{\theta_D} \right) \int_{\frac{\theta_D}{\theta \cdot N_l}}^{\frac{\theta_D}{\theta}} \frac{x^4}{e^x - 1} \left\{ 1 + \frac{1}{2\pi \cdot \left( \frac{\theta N_l}{\theta_D} \right) x} S(x, \theta_e, \theta_D, N_l) \right\} \end{array} \right\} dx \quad (66)$$

where

$$S(x, \theta_e, \theta_D, N_l) = \sum_{\gamma}{}' \frac{\sin \left[ 2\pi \sqrt{j_x^2 + j_y^2 + j_z^2} \cdot \left( \frac{\theta_e N_l}{\theta_D} \right) x \right]}{\sqrt{j_x^2 + j_y^2 + j_z^2}}. \quad (67)$$

This is a generalization to the case of finite cubic crystal, of electron-phonon energy exchange obtained in Eq. (8) for infinite crystal. The biggest difference between electron-phonon energy exchange in finite and infinite crystals occurs for materials with high Debye temperature and low phonons temperature. To demonstrate this, we show the graphs of the ratio of the expression (66) to the corresponding expression obtained by [8]. Let us to denote this relationship as $R(\theta, \theta_D, N_l, \infty)$. The symbol $\infty$ shows that formally the infinite summing has to be performs in Eq. (66), but in practice one may includes only $P$ terms in every sum:

$$R(\theta, \theta_D, N_l, \infty) \equiv \left(\frac{\partial \varepsilon}{\partial t}\right)^{(finite)}_{ph,e} \Bigg/ \left(\frac{\partial \varepsilon}{\partial t}\right)^{(infinite)}_{ph,e} \approx R(\theta, \theta_D, N_l, P) =$$

$$= \frac{\int_{\frac{\theta_D}{\theta \cdot N_l}}^{\frac{\theta_D}{\theta}} \frac{x^3}{e^x - 1} \left\{ x + \sum_{j_x=1}^{P} \sum_{j_y=1}^{P} \sum_{j_z=1}^{P} \frac{8 \sin\left[2\pi \sqrt{j_x^2 + j_y^2 + j_z^2} \cdot \left(\frac{\theta \cdot N_l}{\theta_D}\right) x\right]}{2\pi \cdot \frac{\theta \cdot N_l}{\theta_D} \sqrt{j_x^2 + j_y^2 + j_z^2}} \right\} dx}{\int_0^{\frac{\theta_D}{\theta}} \frac{x^4}{e^x - 1} dx}. \quad (68)$$

The graphs of this expression as a function of the number of atomic layers $N_l$ for $\theta_D = 1440 \cdot k_B$ (Beryllium), $\theta_D = 470 \cdot k_B$ (Iron) $\theta_D = 170 \cdot k_B$ (Gold) are shown on the Fig 2. – Fig 4. The phonon temperature has been chosen as $\theta = 10 \cdot k_B$ and four terms in sums (Eq. (68)) are taken into account.

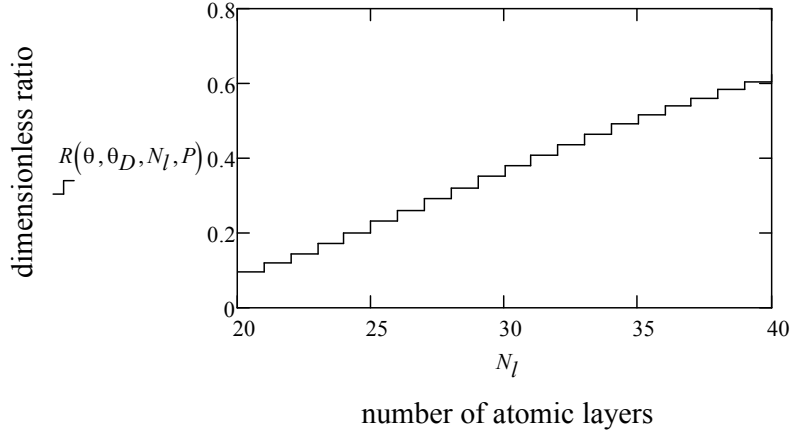

Fig. 2 This chart shows the ratio of the electron-phonon energy exchange in the finite Beryllium ($\theta_D = 1440 \cdot k_B$) domain to the energy exchange in an infinite domain. The number of atomic layers along the edges of the cub shows on the x-axis. Four terms in sums ($P = 4$) in the formula (68) are taken into account. The phonon temperature is $\theta = 10 \cdot k_B$.

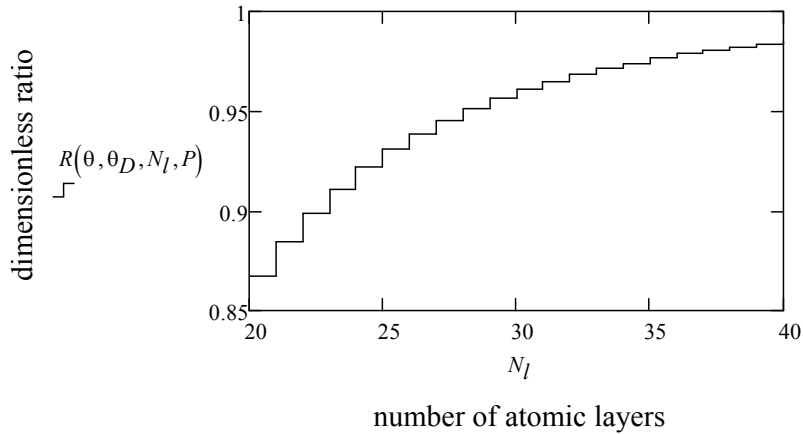

Fig. 3. This chart shows the ratio of the electron-phonon energy exchange in the finite Iron ($\theta_D = 470 \cdot k_B$) domain to the energy exchange in an infinite domain.

The number of atomic layers along the edges of the cub shows on the x-axis. Four terms in sums ($P = 4$) in the formula (68) are taken into account. The phonon temperature is $\theta = 10 \cdot k_B$.

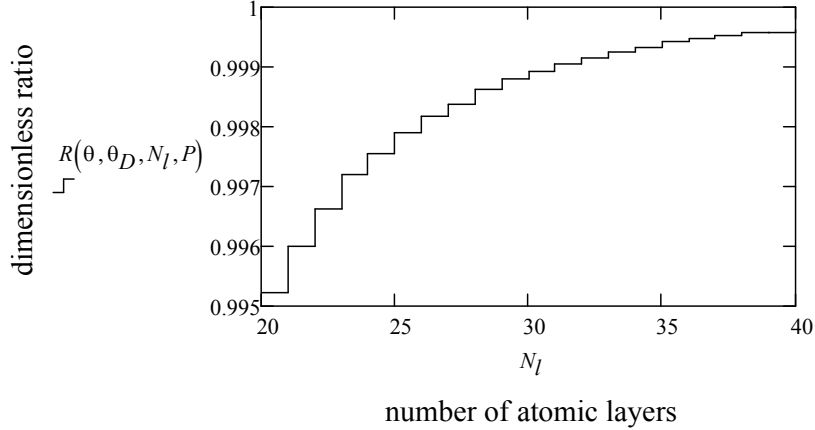

number of atomic layers

Fig. 4. This chart shows the ratio of the electron-phonon energy exchange in the finite Golden ($\theta_D = 170 \cdot k_B$) domain to the energy exchange in an infinite domain. The number of atomic layers along the edges of the cub shows on the x-axis. Four terms in sums ($P = 4$) in the formula (68) are taken into account. The phonon temperature is $\theta = 10 \cdot k_B$.

Fig 2. – Fig 4. show that the discrete phonon spectrum manifested significantly (in the electron-phonon energy exchange) for small particles made from materials with high Debye temperature and at low phonons temperatures. As an example, for Beryllium MN with the size 20 atoms layers ($\sim 45$ Angstroms) at the phonon temperature $\theta = 10 \cdot k_B$ the energy exchange is $\sim 10$ % from infinite Beryllium crystal.

V. RESULTS AND DISCUSSION

In this paper, based on the Bardeen-Shockley deformation potential for the electron-phonon interaction, we developed the quantum kinetic approaches for the finite domain the classical approaches to describe the electron-lattice energy exchange in MN. The basis of quantum-kinetic approach is kinetic Boltzmann equation as the basis for the classic - the equation generating sound vibrations electron moving in the metal.
It was shown that when the cluster size larger than the mean free path of an electron, classic approach gives for the intensity of the electron-lattice energy exchange the same expression as the quantum kinetic approach. However, the classical approach allows a simple generalization to the cluster size smaller than the length of the mean free path of an electron, where the electron motion is ballistic (electron moves freely from wall to wall). In this case, the intensity of the electron-lattice energy exchange begins quasi-vibrationally depends on the size of the cluster, and for certain cluster size, the energy exchange is particularly sensitive to this size.
As for the generalization of the quantum kinetic approach to the transitional regime of the electron (from diffuse to ballistic regime), this problem is more complex and requires special consideration. For the case of low temperatures, within the quantum kinetic approach had the effect of discrete phonon spectrum (caused by the finite size of the cluster) to the electron-lattice energy exchange.

ACKNOWLEDGMENTS

We gratefully acknowledge stimulating discussion with M. Gasik. One of the authors (PT) takes this opportunity of thanking the Fund BC 156 and another author (YB) the Helsinki University of Technology Foundation for the support for this research.